\documentclass[aps,prd,twocolumn,nofootinbib,superscriptaddress]{revtex4}
\usepackage{amssymb,amsmath,graphicx,enumerate}
\usepackage{subfigure}
\usepackage{dcolumn,booktabs}

\usepackage[dvipdfm,colorlinks=true,citecolor=blue,pdfstartview=FitH]{hyperref}

\usepackage{color,framed} 
\definecolor{shadecolor}{rgb}{0.3,0.3,1} 

\def\bea{\begin{equation}}
\def\eea{\end{equation}}
\newcommand{\hxq}{hexaquark ${(\bar{u}(cc))(b(\bar{b}\bar{b}))}$}
\newcommand{\rt}{Regge trajectory}
\newcommand{\rts}{Regge trajectories}
\newcommand{\tr}{trajectory}
\newcommand{\trs}{trajectories}
\newcommand{\bfr}{{\bf r}}
\newcommand{\bfp}{{\bf p}}
\newcommand{\bfpa}{{|\bf p|}}
\newcommand{\gev}{{\rm GeV}}

\newcommand{\cltba}{\bar{3}_c}

\newcommand{\dhbs}{doubly heavy baryons}

\begin{document}
\title{$\lambda$, $\rho$, and $\sigma$ Regge trajectories for the hexaquark ${(\bar{u}(cc))(b(\bar{b}\bar{b}))}$ in the triquark-antitriquark picture}
\author{Xin-Ru Liu}
\email{1170394732@qq.com}
\affiliation{School of Physics and Electronic Engineering, Shanxi Normal University, Taiyuan 030031, China}
\author{Qi Liu}
\email{18803429267@163.com}
\affiliation{School of Physics and Electronic Engineering, Shanxi Normal University, Taiyuan 030031, China}
\author{Jiao-Kai Chen}
\email{chenjk@sxnu.edu.cn, chenjkphy@outlook.com (corresponding author)}
\affiliation{School of Physics and Electronic Engineering, Shanxi Normal University, Taiyuan 030031, China}

\begin{abstract}
We propose Regge trajectory relations for the hexaquark ${(\bar{u}(cc))(b(\bar{b}\bar{b}))}$ by using the Regge trajectory relations for diquarks and triquarks. With these newly derived relations, we investigate five series of hexaquark Regge trajectories: the $\lambda$-, $\rho_1$-, $\rho_2$-, $\sigma_1$-, and $\sigma_2$-trajectories.
We demonstrate that, apart from the simplest $\lambda_1$-trajectories, the $\rho_1$-, $\rho_2$-, $\sigma_1$-, and $\sigma_2$-trajectories cannot be constructed by merely mimicking the meson Regge trajectories, since mesons possess no internal substructures. To derive these trajectories, one must account for the structure and internal substructure of the hexaquark. Without this structural information, the $\rho_1$-, $\rho_2$-, $\sigma_1$-, and $\sigma_2$-trajectories could only be obtained through direct fits to available theoretical predictions or future experimental data.
We demonstrate that the $\rho_1$-, $\rho_2$-, $\sigma_1$-, and $\sigma_2$-trajectories for the hexaquark do not correspond respectively to the Regge trajectories for the triquark, antitriquark, diquark, and antidiquark. Nevertheless, their behaviors match those of the Regge trajectories for the triquark $(\bar{u}(cc))$, the antitriquark $(b(\bar{b}\bar{b}))$, the diquark $(cc)$, and the antidiquark $(\bar{b}\bar{b})$, in that respective order.
Furthermore, we present rough mass estimates for the excited states corresponding to the $\lambda$-, $\rho_1$-, $\rho_2$-, $\sigma_1$-, and $\sigma_2$-trajectories.
\end{abstract}

\maketitle


\section{Introduction}
In 1964, Dyson and Xuong first proposed the possibility of dibaryon hexaquark states \cite{Dyson:1964xwa}. In the original work \cite{Jaffe:1976yi}, Jaffe considered the H dibaryon with quark content $uuddss$, which stimulated extensive theoretical investigations and experimental searches.
In recent years, the hexaquark $d^{\ast}(2380)$ has been observed and its properties have been determined \cite{Bashkanov:2008ih,WASA-at-COSY:2011bjg,WASA-at-COSY:2014dmv,WASA-at-COSY:2014qkg,A2:2019arr}.
Hexaquarks have been studied in various pictures, including the dibaryon picture \cite{Jaffe:1976yi,Chow:1994hg,Bashinsky:1997qv,Huang:2014kja,Yost:1985mj,Farrar:2023wvm}, baryon-antibaryon picture \cite{Abud:2009rk,BES:2003aic,Belle:2004dmq,Belle:2008xmh,Wan:2019ake,Cheng:2022vgy}, diquark-diquark-diquark picture \cite{Wang:2017sto,Azizi:2019xla,Kim:2020rwn}, meson-meson-meson picture \cite{Wang:2020fuh,Valderrama:2018sap,Dias:2017miz,Wu:2021kbu,Liu:2022gxf}, triquark-antitriquark picture \cite{Zhang:2025vqg}, six-quark picture \cite{Lu:2022myk,Pan:2023wrm,Pepin:1998ih,Gordillo:2026usv}, and others.

The {\rt} is one of the effective approaches widely used in the study of hadron spectra
\cite{Burns:2010qq,Regge:1959mz,Chew:1962eu,Nambu:1978bd,Gross:2022hyw,Brodsky:2006uq,Nielsen:2018uyn,
Brau:2000st,Brisudova:1999ut,Guo:2008he,Ebert:2009ub,Irving:1977ea,Collins:1971ff,
Inopin:1999nf,Afonin:2014nya,MartinContreras:2020cyg,Sergeenko:1994ck,Veseli:1996gy,
Inopin:2001ub,Wilczek:2004im,Selem:2006nd,Nambu:1974zg,Kobylinsky:1989dd,Brodsky:2016yod,
Sonnenschein:2018fph,MartinContreras:2023oqs,Roper:2024ovj,Oudichhya:2024ikt,Dey:1994za,
G:2024zkc,Ghaderi:2026yap,Furuichi:1977xh,Gao:2026fuu,Cakir:2026fzd,MonyA:2026mtz,Xu:2025zna,
MartinContreras:2026uyt,He:2026tev,daSilvaJunior:2025wyn,RuizArriola:2025omi,Bernardo:2021yeh,
Chen:2018nnr,Chen:2018bbr,Chen:2018hnx,Kahana:1993yd,Olsson:1994cv,Chen:2022flh}. There are various models to investigate the {\rts}, such as the Nambu string model \cite{Nambu:1974zg,Nambu:1978bd}, the relativistic flux tube model \cite{Burns:2010qq}, the light-front holographic QCD \cite{Brodsky:2016yod,Brodsky:2006uq}, the holographic Ads/QCD context \cite{MartinContreras:2020cyg}, the $\hbar$-expansion technique \cite{Kobylinsky:1989dd}, $\kappa$ deformed algebra approach \cite{Dey:1994za}, the quadratic form of the spinless Salpeter-type equation \cite{Chen:2018hnx}, the relativistic Thompson equation \cite{Kahana:1993yd}, the Dirac equation \cite{Olsson:1994cv}, the spinless Salpeter equation \cite{Brau:2000st,Lucha:1991vn,Chen:2022flh}, and so on. More references can be found in Refs. \cite{Inopin:2001ub,Inopin:1999nf}.
In Ref. \cite{Inopin:1999nf}, the authors scrutinize the hadron Regge trajectories in the framework of the string and potential models.
In the present work, we do not analyze string states, instead, we consider composite states of quarks and antiquarks subjected to a potential that includes a linear term in the separation between particles.
We employ the spinless Salpeter equation to derive the {\rt} relations \cite{Chen:2022flh} and subsequently construct the {\rt} relations for the quintuply heavy {\hxq}. Five families of {\rts} for the quintuply heavy {\hxq} are systematically investigated.

Studies on hexaquark {\rts} remain scarce \cite{G:2024zkc,Ghaderi:2026yap,Furuichi:1977xh} owing to their complexity, and all prior works only addressed $\lambda$-{\trs}.
We not only explore the $\lambda$-{\trs} for the {\hxq} but also construct the $\rho_1$-, $\rho_2$, $\sigma_1$- and $\sigma_2$-{\trs} for the quintuply heavy hexaquark, on the basis of established diquark {\rt} relations \cite{Feng:2023txx} and triquark {\rts} \cite{Song:2024bkj,Liu:2026btq}.
We present the behaviors of these five families of {\rts} and provide rough mass predictions for the excited states of all five families.

The paper is organized as follows: Sec. \ref{sec:rgr} establishes the {\rt} relations for the {\hxq}. Section \ref{sec:rts} elaborates five series of {\rts} and carries out approximate mass evaluations of all excited states. The conclusions are presented in Sec. \ref{sec:conc}.

\section{{\rt} relations for the hexaquark ${(\bar{u}(cc))(b(\bar{b}\bar{b}))}$}\label{sec:rgr}
In this section, by utilizing the diquark {\rts} \cite{Feng:2023txx} and the triquark {\rts} \cite{Song:2024bkj,Liu:2026btq}, we propose the hexaquark {\rts}, which can be employed to discuss $\lambda$-, $\rho$-, and $\sigma$-trajectories.

\subsection{Preliminary}\label{subsec:prelim}

\begin{figure}[!phtb]
\centering
\includegraphics[width=0.26\textheight]{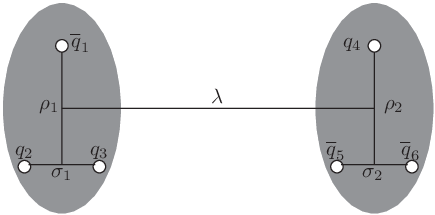}
\caption{Schematic diagram of a hexaquark in the triquark-antitriquark picture. The left gray part represents the triquark $(\bar{q}_1(q_2q_3))$, composed of an antiquark $\bar{q}_1$ and a diquark $(q_2q_3)$. The right gray part represents an antitriquark $(q_4(\bar{q}_5\bar{q}_6))$, composed of a quark ($q_4$) and an antidiquark $(\bar{q}_5\bar{q}_6)$. The circles denote quarks and antiquarks. The lines represent separations between two constituents or two clusters. $\lambda$ separates the triquark and the antitriquark. $\rho_1$ separates the antiquark $\bar{q}_1$ and the diquark $(q_2q_3)$ inside the triquark $(\bar{q}_1(q_2q_3))$. $\rho_2$ separates the quark ${q}_4$ and the antidiquark $(\bar{q}_5\bar{q}_6)$ in the antitriquark. $\sigma_1$ separates $q_2$ and $q_3$ in the diquark. $\sigma_2$ separates $\bar{q}_5$ and $\bar{q}_6$ in the antidiquark. }\label{fig:pr}
\end{figure}

Figure \ref{fig:pr} shows a schematic diagram of a hexaquark in the potential model.
In the triquark-antitriquark picture, a hexaquark consists of one triquark and one antitriquark. The lines represent separations between two constituents or two clusters. $\lambda$ corresponds to the separation between the triquark and the antitriquark.
$\rho_1$ separates the antiquark $\bar{q}_1$ and the diquark $(q_2q_2)$ inside the triquark $(\bar{q}_1(q_2q_3))$. Similarly, $\rho_2$ separates the quark ${q}_4$ and the antidiquark $(\bar{q}_5\bar{q}_6)$ within the antitriquark $(q_4(\bar{q}_5\bar{q}_6))$ subsystem.
$\sigma_1$ corresponds to the separation between quark $q_2$ and quark $q_3$, while $\sigma_2$ corresponds to the separation between the antiquark $\bar{q}_5$ and antiquark $\bar{q}_6$.
Five excited modes arise from these relative coordinates: the $\sigma_1$-mode involves radial and orbital excitations within the diquark $(q_2q_3)$, the $\sigma_2$-mode corresponds to radial and orbital excitations inside the antidiquark $(\bar{q}_5\bar{q}_6)$, the $\rho_1$-mode accounts for radial and orbital excitations between the antiquark $\bar{q}_1$ and diquark $(q_2q_3)$ insde the triquark, the $\rho_2$-mode involves excitations between quark $q_4$ and the antidiquark $(\bar{q}_5\bar{q}_6)$ within the antitriquark, and the $\lambda$-mode involves radial and orbital excitations arising from the relative motion of the entire triquark and antitriquark subsystems. Accordingly, five series of {\rts} can be classified from these modes: two series of $\sigma$-{\trs}, two series of $\rho$-{\trs}, and one series of $\lambda$-{\trs}.

A diquark $(q_2q_3)$ couples to two irreducible color representations via the product $3_c\otimes3_c=\cltba\oplus{6}_c$. The $\bar{3}_c$ channel supports an attractive interaction, whereas the $6_c$ representation yields repulsive interquark forces within the diquark pair.
Some previous analyses incorporate both $\cltba$ and ${6}_c$ color channels (e.g.,Refs. \cite{Maiani:2019lpu,Berwein:2024ztx}), whereas other studies employ only the $\cltba$ channel (e.g., Refs. \cite{Giron:2021fnl,Brodsky:2014xia,Galkin:2023wox}).
Consistent with the treatments in Refs. \cite{Giron:2021fnl,Brodsky:2014xia,Galkin:2023wox}, we adopt only the $\bar{3}_c$ diquark configuration throughout this paper. By the same logic, the triquark $(\bar{q}_1(q_2q_3))$ studied here forms a ${3}_c$ bound state assembled from a $\cltba$ diquark $(q_2q_3)$ and a $\cltba$ antiquark $\bar{q}_1$. In the triquark-antitriquark model, the color-singlet hexaquarks under consideration are constructed from a ${3}_c$ triquark and a $\bar{3}_c$ antitriquark.

\begin{widetext}
In the triquark-antitriquark picture, the state of a hexaquark is denoted by
\bea\label{tetnot}
\left(\left(\bar{q}_1(q_2q_3)^{{\bar{3}_c}}_{n_3^{2s_{3}+1}l_{3j_3}}\right)
^{{{3}_c}}_{n_1^{2s_{1}+1}l_{1j_1}}
\left({q}_4(\bar{q}_5\bar{q}_6)^{{{3}_c}}_{n_4^{2s_4+1}l_{4j_4}}\right)
^{{\bar{3}_c}}_{n_2^{2s_{2}+1}l_{2j_2}}
\right)^{1_c}_{N^{2S+1}L_{J}},
\eea
\end{widetext}
where $1_c$ represents the color singlet state of the hexaquark. (The superscript $1_c$ is often omitted, as the observed hexaquarks are colorless.)
The notation defined in Eq. (\ref{tetnot}) can be abbreviated as $|n_1^{2s_1+1}l_{1j_1},n_2^{2s_2+1}l_{2j_2},n_3^{2s_3+1}l_{3j_3},n_4^{2s_4+1}l_{4j_4},N^{2S+1}L_{J}\rangle$.
The diquark $(q_2q_3)$ is either $\{q_2q_3\}$ or $[q_2q_3]$, where $\{q_2q_3\}$ and $[q_2q_3]$ represent the permutation symmetric and antisymmetric flavor wave functions, respectively. $N=N_{r}+1$, where $N_{r}=0,\,1,\,\cdots$. $n_{1,2,3,4}=n_{r_{1,2,3,4}}+1$, where $n_{r_{1,2,3,4}}=0,\,1,\,\cdots$. $N_{r}$, $n_{r_1}$, $n_{r_2}$, $n_{r_3}$, and $n_{r_4}$ are the radial quantum numbers of the hexaquark, triquark $(\bar{q}_1(q_2q_3))$, antitriquark $(q_4(\bar{q}_5\bar{q}_6))$, diquark $(q_2q_3)$, and antidiquark $(\bar{q}_5\bar{q}_6)$, respectively. The angular momentum coupling chain reads as follows:
\begin{align}
\vec{J}=&\vec{L}+\vec{S},\; \vec{S}=\vec{j}_1+\vec{j}_2, \nonumber\\ \vec{j}_1=&\vec{s}_{1}+\vec{l}_1,\; \vec{s}_{1}=\vec{s}_{\bar{q}_1}+\vec{j}_{3}, \nonumber 
\end{align}
\begin{align}
 \vec{j}_2=&\vec{s}_2+\vec{l}_2,\; \vec{s}_2=\vec{s}_{{q}_4}+\vec{j}_4, \nonumber \\
 \vec{j}_3=&\vec{s}_{3}+\vec{l}_3,\; \vec{s}_{3}=\vec{s}_{q_2}+\vec{s}_{q_3}, \nonumber\\
\vec{j}_4=&\vec{s}_4+\vec{l}_4,\; \vec{s}_4=\vec{s}_{\bar{q}_5}+\vec{s}_{\bar{q}_6}.
\end{align}
$\vec{J}$, $\vec{j}_1$, $\vec{j}_2$, $\vec{j}_3$,  and $\vec{j}_4$ are the spins of the hexaquark, triquark, antitriquark, diquark, and antidiquark, respectively. $L$, $l_1$, $l_{2}$, $l_{3}$, and $l_4$ are the orbital quantum numbers of the hexaquark, triquark, antitriquark, diquark, and antidiquark, respectively. $\vec{s}_{1}$ is the summed spin of antiquark $\bar{q}_1$ and diquark $(q_2q_3)$, $\vec{s}_{2}$ is the summed spin of quark ${q}_4$ and antidiquark $(\bar{q}_5\bar{q}_6)$, $\vec{s}_{3}$ is the summed spin of quarks in the diquark, $\vec{s}_{4}$ is the summed spin of antiquarks in the antidiquark, and $\vec{S}$ is the summed spin of triquark and antitriquark.

In the triquark-antitriquark picture, the {\hxq} has four configurations: $(\bar{u}(cc))(b(\bar{b}\bar{b}))$, $(\bar{b}(cc))(b(\bar{u}\bar{b}))$, $(\bar{u}(bc))(c(\bar{b}\bar{b}))$, and $(\bar{b}(bc))(c(\bar{u}\bar{b}))$. Due to mode mixings in the latter three configurations, which renders their {\rt} descriptions complicated, we exclude these three cases from the scope of this study.

\subsection{Spinless Salpeter equation}
The spinless Salpeter equation \cite{Godfrey:1985xj,Capstick:1986ter,Ferretti:2019zyh,Bedolla:2019zwg,Durand:1981my,Durand:1983bg,Lichtenberg:1982jp,Jacobs:1986gv} takes the form
\begin{eqnarray}\label{qsse}
M\Psi_{d,t,h}({\bfr})=\left(\omega_1+\omega_2\right)\Psi_{d,t,h}({\bfr})+V_{d,t,h}\Psi_{d,t,h}({\bfr}),
\end{eqnarray}
where $M$ denotes the bound state mass (diquark, triquark, and hexaquark). $\Psi_{d,t,h}({\bfr})$ are the diquark, triquark, and hexaquark wave functions, respectively. $V_{d,t,h}$ stands for the diquark, triquark, and hexaquark potentials, respectively (see Eq. (\ref{potv})). $\omega_1$ represents the relativistic energy of constituent $1$ (quark, antiquark, diquark, antidiquark, and triquark), and $\omega_2$ corresponds to the relativistic energy of constituent $2$ (quark, antiquark, diquark, antidiquark, and antitriquark),
\bea\label{omega}
\omega_i=\sqrt{m_i^2+{\bf p}^2}=\sqrt{m_i^2-\Delta}\;\; (i=1,2).
\eea
$m_1$ and $m_2$ are the effective masses of constituent $1$ and $2$, respectively.

Following Refs. \cite{Ferretti:2019zyh,Bedolla:2019zwg,Ferretti:2011zz,Eichten:1974af}, we employ the potential
\begin{align}\label{potv}
V_{d,t,h}&=-\frac{3}{4}\left[V_c+{\sigma}r+C\right]
\left({\bf{F}_i}\cdot{\bf{F}_j}\right)_{d,t,h},
\end{align}
where $V_c\propto{1/r}$ is a color Coulomb potential or a Coulomb-like potential due to one-gluon exchange. $\sigma$ is the string tension. $C$ is a fundamental parameter \cite{Gromes:1981cb,Lucha:1991vn}. The part in the bracket is the Cornell potential \cite{Eichten:1974af}. ${\bf{F}_i}\cdot{\bf{F}_j}$ is the color-Casimir,
\bea\label{mrcc}
\langle{(\bf{F}_i}\cdot{\bf{F}_j})_{d,t}\rangle=-\frac{2}{3},\quad
\langle{(\bf{F}_i}\cdot{\bf{F}_j})_{h}\rangle=-\frac{4}{3}.
\eea

\subsection{{\rt} relations for various systems}
For heavy-heavy systems with $m_{1},m_2{\gg}{\bfpa}$, Eq. (\ref{qsse}) simplifies to
\begin{eqnarray}\label{qssenrr}
M\Psi_{d,t,h}({\bfr})=\left[(m_1+m_2)+\frac{{\bfp}^2}{2\mu}+V_{d,t,h}\right]\Psi_{d,t,h}({\bfr}),
\end{eqnarray}
where
\bea\label{rdmu}
\mu=\frac{m_1m_2}{m_1+m_2}.
\eea
Applying the Bohr-Sommerfeld quantization approach \cite{Brau:2000st} together with Eqs. (\ref{potv}) and (\ref{qssenrr}), we obtain the parametrized relation  \cite{Chen:2022flh,Chen:2021kfw}
\bea\label{massform}
M=m_R+\beta_x(x+c_{0x})^{2/3}\,\,(x=l,\,n_r,\,L,\,N_r).
\eea
Here, the {\tr} coefficients are defined as
\bea\label{parabm}
\beta_x=c_{fx}c_xc_c,\quad m_R=m_1+m_2+C',
\eea
where
\begin{align}\label{cprime}
C'=\left\{\begin{array}{ll}
C/2 &\text{for diquarks and triquarks}, \\
C &\text{for hexaquarks}.
\end{array}\right. \nonumber\\
\sigma'=\left\{\begin{array}{ll}
\sigma/2 &\text{for diquarks and triquarks}, \\
\sigma &\text{for hexaquarks}.
\end{array}\right.
\end{align}
$c_{x}$ and $c_c$ are
\bea\label{cxcons}
c_c=\left(\frac{\sigma'^2}{\mu}\right)^{1/3},\quad c_{l,L}=\frac{3}{2},\quad c_{n_r,N_r}=\frac{\left(3\pi\right)^{2/3}}{2}.
\eea
$c_{fx}$ theoretically equals unity, yet it is treated as a fitting parameter in practice.
In Eq. (\ref{massform}), $m_1$, $m_2$, $c_x$ and $\sigma$ remain universal constants for heavy-heavy systems. $c_{0x}$ varies across different {\rts}.
The {\rt} relation in Eq. (\ref{massform}) for heavy-heavy systems obtained from the potential model agrees with the results obtained from the relativistic hadron string model (see Eq. (17) in Ref. \cite{Burns:2010qq}) and the holography inspired stringy hadron model (see Eq. (3.6) in Ref. \cite{Sonnenschein:2018fph}), among others.

For heavy-light systems ($m_1\to\infty$ and $m_2\to0$), Eq. (\ref{qsse}) simplifies to
\begin{eqnarray}\label{qssenr}
M\Psi_{d,t,h}({\bfr})=\left[m_1+{\bfpa}+V_{d,t,h}\right]\Psi_{d,t,h}({\bfr}).
\end{eqnarray}
Upon applying the Bohr-Sommerfeld quantization approach \cite{Brau:2000st} to Eq. (\ref{qssenr}), we obtain the parametrized formula \cite{Chen:2022flh,Chen:2021kfw}
\bea\label{rtmeson}
M=m_R+\beta_x\sqrt{x+c_{0x}}\;(x=l,\,n_r,\,L,\,N_r).
\eea
$\beta_x$ is given in Eq. (\ref{parabm}), while the constants become
\bea\label{cxconshl}
c_{c}=\sqrt{\sigma'},\quad c_{l,L}=2,\quad c_{n_r,N_r}=\sqrt{2\pi}.
\eea
For heavy-light systems, the common choice of $m_R$ is \cite{Selem:2006nd,Chen:2021kfw,Veseli:1996gy}
\bea\label{mrm1}
m_R=m_1.
\eea
Equation (\ref{rtmeson}) with (\ref{mrm1}) is consistent with the results obtained from the loaded flux tube model (see Eq. (13) in Ref. \cite{Selem:2006nd} and Eq. (2.13) in Ref. \cite{Wilczek:2004im}).

The conventional {\rt}, Eq. (\ref{rtmeson}) with (\ref{mrm1}), is obtained in the limit $m_1\to\infty$ and $m_2\to0$.
In Ref. \cite{Chen:2023cws}, we propose two modified formulas that explicitly incorporate the mass of light constituents, allowing an unified description of both heavy-light mesons and heavy-light diquarks.
One is Eq. (\ref{rtmeson}) with $m_R$ given in (\ref{parabm}), where $m_2$ is the light constituent's mass. The other reads
\bea\label{mrtf}
M=m_R+\sqrt{\beta_x^2(x+c_{0x})+\kappa_{x}m^{3/2}_2(x+c_{0x})^{1/4}}
\eea
for $m_2{\ll}M$, where
\bea\label{mrfp}
m_R=m_1+C',\quad \kappa_x=\frac{4}{3}\sqrt{{\pi}\beta_x},
\eea
where $\beta_x$ is given in (\ref{parabm}).
Equation (\ref{rtmeson}) with (\ref{parabm}) extends the formula
\bea\label{afoninequ}
M=m_1+m_2+\sqrt{a(n_r+{\alpha}l+b)}
\eea
from Ref. \cite{Afonin:2014nya} and the formula
\bea\label{rmfnpb}
(M-m_1-m_2-C)^2=\alpha_x(x+c_0)^{\gamma}
\eea
from Ref. \cite{Chen:2022flh} whereas Eq. (\ref{mrtf}) with (\ref{mrfp}) is based on the results of \cite{Selem:2006nd,Sonnenschein:2018fph}.
As $m_2=0$, these two modified formulas, Eq. (\ref{rtmeson}) with (\ref{parabm}) and Eq. (\ref{mrtf}) with (\ref{mrfp}), become identical. As $m_2=0$ and $C$ is neglected, these two modified formulas reduce to the usual {\rt} formula for heavy-light systems, i.e., Eq. (\ref{rtmeson}) with (\ref{mrm1}).
Although they give different behavior of $m_2$, Eq. (\ref{rtmeson}) with (\ref{parabm}) and Eq. (\ref{mrtf}) with (\ref{mrfp}) produce consistent results for $l,\,n_r<10$ and have the same behavior $M{\sim}x^{1/2}$ \cite{Chen:2023cws}.

For light systems $(m_1,\,m_2\to0)$, Eq. (\ref{qsse}) simplifies to
\begin{eqnarray}\label{qsslight}
M\Psi_{d,t,h}({\bfr})=\Large[\,2|\bfp|+V_{d,t,h}\,\Large]\Psi_{d,t,h}({\bfr}).
\end{eqnarray}
Applying the Bohr-Sommerfeld quantization approach \cite{Brau:2000st} to the ultrarelativistic limit in Eq. (\ref{qsslight}), we obtain the parametrized formula \cite{Chen:2022flh,Chen:2021kfw}
\bea\label{ultrts}
M=\beta_x\sqrt{x+c_{0x}}\;(x=l,\,n_r,\,L,\,N_r).
\eea
$\beta_x$ is given in Eq. (\ref{parabm}), while the constants become
\bea\label{ultconshx}
c_{c}=\sqrt{\sigma'},\quad c_{l,L}=2\sqrt{2},\quad c_{n_r,N_r}=2\sqrt{\pi}.
\eea
Evidently, Eq. (\ref{ultrts}) is consistent with the familiar form $\alpha'M^2=J$.
Based on Eqs. (\ref{afoninequ}) and (\ref{rmfnpb}), we obtain a modified {\rt} relation for light systems in Ref. \cite{Chen:2023ngj},
\bea\label{ultrartf}
M=m_R+\beta_x\sqrt{x+c_{0x}},
\eea
where $m_R$ is in Eq. (\ref{parabm}) with $m_{1,2}$ being the masses of the light constituents.

\begin{table}[!phtb]
\caption{The coefficients for heavy-heavy systems (HHSs), heavy-light systems (HLSs), and light-light systems (LLSs) \cite{Xie:2024dfe}.}  \label{tab:eparam}
\centering
\begin{tabular*}{0.47\textwidth}{@{\extracolsep{\fill}}cccc@{}}
\hline\hline
                   & HHSs &  HLSs & LLSs  \\
\hline
$\nu$    & $2/3$ & $1/2$  & $1/2$  \\
$c_c$    & $\left({\sigma'^2}/{\mu}\right)^{1/3}$    & $\sqrt{\sigma'}$  & $\sqrt{\sigma'}$ \\
$c_{l,L}$    & $3/2$ & $2$  & $2\sqrt{2}$  \\
$c_{n_r,N_r}$ & ${\left(3\pi\right)^{2/3}}/{2}$      & $\sqrt{2\pi}$ & $2\sqrt{\pi}$   \\
\hline
\hline
\end{tabular*}
\end{table}

By unifying the descriptions for heavy-heavy and heavy-light systems via combining Eqs. (\ref{massform}), (\ref{rtmeson}), (\ref{ultrartf}), and (\ref{parabm}), we arrive at a general form of the {\rts} \cite{Chen:2022flh,Xie:2024lfo}
\begin{align}\label{massfinal}
M=&m_R+\beta_x(x+c_{0x})^{\nu}\,\,(x=l,\,n_r,\,L,\,N_r),\nonumber\\
m_R=&m_1+m_2+C',\quad \beta_x=c_{fx}c_xc_{c},
\end{align}
where ${\nu}$, $c_x$ and $c_{c}$ are listed in Table \ref{tab:eparam}. The parameter $c_{fx}$ is theoretically equal to unity but treated as a free fitting parameter in practice. $c_{0x}$ varies for different {\rts}. Equation (\ref{massfinal}) can be employed to discuss various systems covering both heavy-heavy, heavy-light, and light-light systems: diquarks, mesons, baryons, triquarks, tetraquarks, and pentaquarks \cite{Chen:2023djq,Chen:2023ngj,Chen:2023web,Chen:2025fyh,Song:2025cla,Song:2024bkj}.

It is worth emphasizing that the universal form in Eq. (\ref{massfinal}) remains provisional. Multiple alternative schemes have been proposed to incorporate masses of light constituents, so further theoretical calculations and experimental measurements are required to distinguish a better one. Furthermore, the fitted parameter set holds universal validity for both heavy-heavy and heavy-light systems \cite{Chen:2023cws,Feng:2023txx}. By contrast, the parameter values should be adjusted for light systems \cite{Chen:2023ngj} to achieve agreeable results.

\subsection{{\rt} relations for the {\hxq} }

The {\hxq} consists of one heavy triquark and one heavy antitriquark, therefore, it is a heavy-heavy system for the $\lambda$-mode. The triquark $(\bar{u}(cc))$ contains one light antiquark and one doubly heavy diquark, hence, it is a heavy-light system for the $\rho_1$-mode. For the $\rho_2$-mode, the antitriquark $(b(\bar{b}\bar{b}))$ is evidently a heavy-heavy system. Since both the diquark inside the triquark and the antidiquark inside the antitriquark belong to heavy-heavy systems, the {\hxq} behaves as a heavy-heavy system for the $\sigma_1$-mode and $\sigma_2$-mode. Using Eqs. (\ref{massform}), (\ref{parabm}), (\ref{cprime}), (\ref{cxcons}), (\ref{rtmeson}), and (\ref{cxconshl}), or Eq. (\ref{massfinal}) and coefficients listed in Table \ref{tab:eparam}, we have the {\rt} relations for the {\hxq}
\begin{align}\label{t2q}
M&=m_{R_{\lambda}}+\beta_{x_{\lambda}}(x_{\lambda}+c_{0x_{\lambda}})^{2/3}\;
(x_{\lambda}=L,\,N_r),\nonumber\\
M_{t_1}&=m_{R_{\rho_1}}+\beta_{x_{\rho_1}}\sqrt{x_{\rho_1}+c_{0x_{\rho_1}}}\;
(x_{\rho_1}=l_1,\,n_{r_1}),\nonumber\\
M_{t_2}&=m_{R_{\rho_2}}+\beta_{x_{\rho_2}}(x_{\rho_2}+c_{0x_{\rho_2}})^{2/3}\;
(x_{\rho_2}=l_2,\,n_{r_2}),\nonumber\\
M_{d_1}&=m_{R_{\sigma_1}}+\beta_{x_{\sigma_1}}(x_{\sigma_1}+c_{0x_{\sigma_1}})^{2/3}\;
(x_{\sigma_1}=l_3,\,n_{r_3}),\nonumber\\
M_{d_2}&=m_{R_{\sigma_2}}+\beta_{x_{\sigma_2}}(x_{\sigma_2}+c_{0x_{\sigma_2}})^{2/3}\;
(x_{\sigma_2}=l_4,\,n_{r_4}),
\end{align}
where
\begin{align}
m_{R_{\lambda}}&=M_{t_1}+M_{t_2}+C,\nonumber\\
m_{R_{\rho_1}}&=M_{d_1}+m_{u}+C/2,\;
m_{R_{\rho_2}}=M_{d_2}+m_{b}+C/2,\nonumber\\
m_{R_{\sigma_1}}&=2m_{c}+C/2,\;
m_{R_{\sigma_2}}=2m_{b}+C/2,\nonumber \\
\beta_{L}&=\frac{3}{2}\left(\frac{\sigma^2}{\mu_{\lambda}}\right)^{1/3}c_{fL},\; \beta_{N_r}=\frac{(3\pi)^{2/3}}{2}\left(\frac{\sigma^2}{\mu_{\lambda}}\right)^{1/3}
c_{fN_r},\nonumber \\
\beta_{l_1}&=\sqrt{2\sigma}c_{fl_1},\; \beta_{n_{r_1}}=\sqrt{\pi\sigma}c_{fn_{r_1}},\;
\mu_{\lambda}=\frac{M_{t_1}M_{t_2}}{M_{t_1}+M_{t_2}},\nonumber 
\end{align}
\begin{align}
\beta_{l_2}&=\frac{3}{2}\left(\frac{\sigma^2}{4\mu_{\rho_2}}\right)^{1/3}c_{fl_2},\;
\mu_{\rho_2}=\frac{M_{d_2}m_{b}}{M_{d_2}+m_{b}},\nonumber\\ \beta_{n_{r_2}}&=\frac{(3\pi)^{2/3}}{2}\left(\frac{\sigma^2}{4\mu_{\rho_2}}\right)^{1/3}
c_{fn_{r_2}},\nonumber\\
\beta_{l_3}&=\frac{3}{2}\left(\frac{\sigma^2}{4\mu_{\sigma_1}}\right)^{1/3}c_{fl_3},\; \mu_{\sigma_1}=\frac{m_c}{2},\nonumber \\
\beta_{n_{r_3}}&=\frac{(3\pi)^{2/3}}{2}\left(\frac{\sigma^2}{4\mu_{\sigma_1}}\right)^{1/3}
c_{fn_{r_3}},\nonumber\\
\beta_{l_4}&=\frac{3}{2}\left(\frac{\sigma^2}{4\mu_{\sigma_2}}\right)^{1/3}c_{fl_4},\; \mu_{\sigma_2}=\frac{m_b}{2}, \nonumber\\ \beta_{n_{r_4}}&=\frac{(3\pi)^{2/3}}{2}\left(\frac{\sigma^2}{4\mu_{\sigma_2}}\right)^{1/3}
c_{fn_{r_4}}.\label{pa2qQ}
\end{align}
In Eq. (\ref{t2q}), $M$, $M_{t_1}$, $M_{t_2}$, $M_{d_1}$, $M_{d_2}$, $m_c$, $m_u$, and $m_b$ are the masses of the hexaquark, triquark, antitriquark, diquark, antidiquark, charm quark, up quark, and bottom quark, respectively. The {\rt} relations for the {\hxq} are given in Eqs. (\ref{t2q}) and (\ref{pa2qQ}), which can be employed to crudely estimate masses of {\hxq}.

According to Eqs. (\ref{t2q}) and (\ref{pa2qQ}), we have
\bea
M=M_{t_1}+M_{t_2}+C+\beta_{x_{\lambda}}(x_{\lambda}+c_{0x_{\lambda}})^{2/3}
\eea
when the triquark and antitriquark are regarded as constituents and the structures of the triquark and antitriquark are not considered. Correspondingly, we have the binding energy, $\epsilon=C+\beta_{x_{\lambda}}(x_{\lambda}+c_{0x_{\lambda}})^{2/3}$. When triquark, antitriquark, diquark and antidiquark are treated as bound states, we have
\begin{align}\label{combrt}
M=&2m_{c}+3m_{b}+m_{u}+3C+\beta_{x_{\lambda}}(x_{\lambda}+c_{0x_{\lambda}})^{2/3}\nonumber\\
&+\beta_{x_{\rho_1}}\sqrt{x_{\rho_1}+c_{0x_{\rho_1}}}
+\beta_{x_{\rho_2}}(x_{\rho_2}+c_{0x_{\rho_2}})^{2/3} \nonumber\\
&+\beta_{x_{\sigma_1}}(x_{\sigma_1}+c_{0x_{\sigma_1}})^{2/3}
+\beta_{x_{\sigma_2}}(x_{\sigma_2}+c_{0x_{\sigma_2}})^{2/3}
\end{align}
from Eqs. (\ref{t2q}) and (\ref{pa2qQ}). Accordingly, the corresponding binding energy is given by $\epsilon=3C+\beta_{x_{\lambda}}(x_{\lambda}+c_{0x_{\lambda}})^{2/3}
+\beta_{x_{\rho_1}}\sqrt{x_{\rho_1}+c_{0x_{\rho_1}}}
+\beta_{x_{\rho_2}}(x_{\rho_2}+c_{0x_{\rho_2}})^{2/3}
+\beta_{x_{\sigma_1}}(x_{\sigma_1}+c_{0x_{\sigma_1}})^{2/3}
+\beta_{x_{\sigma_2}}(x_{\sigma_2}+c_{0x_{\sigma_2}})^{2/3}$.
We can see from Eq. (\ref{combrt}) that the {\hxq} possesses five series of {\rts}: the $\lambda$-trajectories (with $x_{\rho_1}$, $x_{\rho_2}$, $x_{\sigma_1}$, and $x_{\sigma_2}$ fixed), the $\rho_1$-trajectories (with $x_{\lambda}$, $x_{\rho_2}$, $x_{\sigma_1}$, and $x_{\sigma_2}$ fixed), the $\rho_2$-trajectories (with $x_{\lambda}$, $x_{\rho_1}$, $x_{\sigma_1}$, and $x_{\sigma_2}$ fixed), the $\sigma_1$-trajectories (with $x_{\lambda}$, $x_{\rho_1}$, $x_{\rho_2}$, and $x_{\sigma_2}$ fixed), and $\sigma_2$-trajectories (with $x_{\lambda}$, $x_{\rho_1}$, $x_{\rho_2}$, and $x_{\sigma_1}$ fixed).

The {\rts} obtained from Eqs. (\ref{t2q}) and (\ref{pa2qQ}) or from Eqs. (\ref{combrt}) and (\ref{pa2qQ}) are defined as their complete forms. The obtained constant and the mode under consideration are referred to the main part of the {\rts}.
For the $\sigma_2$-{\trs}, $\beta_{x_{\lambda}}(x_{\lambda}+c_{0x_{\lambda}})^{2/3}$ and $\beta_{x_{\rho_2}}(x_{\rho_2}+c_{0x_{\rho_2}})^{2/3}$ become functions of $x_{\sigma_2}$ through the dependence of $\beta_{x_{\lambda}}$ and $\beta_{x_{\rho_2}}$, respectively. Therefore, the main part is
\bea
\widetilde{m}_R+\beta_{x_{\sigma_2}}(x_{\sigma_2}+c_{0x_{\sigma_2}})^{2/3},
\eea
where
\begin{align}
\widetilde{m}_R=&2m_{c}+3m_{b}+m_{u}+3C+
\beta_{x_{\rho_1}}\sqrt{x_{\rho_1}+c_{0x_{\rho_1}}} \nonumber\\
&+\beta_{x_{\sigma_1}}(x_{\sigma_1}+c_{0x_{\sigma_1}})^{2/3}.
\end{align}
For the $\sigma_2$-{\trs}, the main parts of the {\rts} are not equal to the complete forms of the {\rts}. $\sigma_1$-{\trs}, $\rho_1$-{\trs}, and $\rho_2$-{\trs} are similar to $\sigma_2$-{\trs}.
However, for the $\lambda$-{\trs}, $\beta_{x_{\rho_1}}\sqrt{x_{\rho_1}+c_{0x_{\rho_1}}}
+\beta_{x_{\rho_2}}(x_{\rho_2}+c_{0x_{\rho_2}})^{2/3}
+\beta_{x_{\sigma_1}}(x_{\sigma_1}+c_{0x_{\sigma_1}})^{2/3}
+\beta_{x_{\sigma_2}}(x_{\sigma_2}+c_{0x_{\sigma_2}})^{2/3}$ becomes constant. Accordingly, the main part is
\bea
\widetilde{m}_R+\beta_{x_{\lambda}}(x_{\lambda}+c_{0x_{\lambda}})^{2/3},
\eea
where
\begin{align}
\widetilde{m}_R=&2m_{c}+3m_{b}+m_{u}+3C+
\beta_{x_{\rho_1}}\sqrt{x_{\rho_1}+c_{0x_{\rho_1}}}\nonumber\\
&+\beta_{x_{\rho_2}}(x_{\rho_2}+c_{0x_{\rho_2}})^{2/3}
+\beta_{x_{\sigma_1}}(x_{\sigma_1}+c_{0x_{\sigma_1}})^{2/3}\nonumber\\
&+\beta_{x_{\sigma_2}}(x_{\sigma_2}+c_{0x_{\sigma_2}})^{2/3}.
\end{align}
Consequently, for the $\lambda$-{\trs}, the main parts are identical to their complete forms. The $\lambda$-{\trs} possess the simplest form among the five series of {\trs}.

\section{{\rts} for the hexaquark ${(\bar{u}(cc))(b(\bar{b}\bar{b}))}$}\label{sec:rts}
In this section, five series of {\rts} for the {\hxq} are discussed using Eqs. (\ref{t2q}) and (\ref{pa2qQ}) or Eqs. (\ref{combrt}) and (\ref{pa2qQ}). Simultaneously, the masses of the {\hxq} are crudely estimated.

\subsection{Parameters}

The parameter values are listed in Table \ref{tab:parmv}. The values of $m_u$, $m_b$, $m_c$, $\sigma$, and $C$ are directly taken from Ref. \cite{Faustov:2021hjs}. $c_{fx}$ and $c_{0x}$ for the $\sigma$-modes are from Ref. \cite{Feng:2023txx}.
The {\rt} parameters corresponding to the $\lambda$- and $\rho_2$-modes are calculated via the following relations \cite{Xie:2024dfe}: \footnote{In our previous work \cite{Xie:2024dfe,Song:2024bkj,Song:2025cla,Chen:2025fyh,Liu:2026vpi}, $m_0$ [$m_0=1$ {\gev}] was not included in the relations $c_{fL}=1.116 + 0.013 \mu_{\lambda}$, $c_{0L}=0.540- 0.141\mu_{\lambda}$, $c_{fN_{r}}=1.008 + 0.008\mu_{\lambda}$, and $c_{0N_{r}}=0.334 - 0.087\mu_{\lambda}$. In the present work, we correct this and introduce $m_0$ into Eq. (\ref{fitcfxnr}).}
\begin{align}
c_{fL,fl_2}=&1.116 + 0.013\frac{\mu_{\lambda,\rho_2}}{m_0},\nonumber 
\end{align}
\begin{align}
c_{0L,0l_2}=&0.540- 0.141\frac{\mu_{\lambda,\rho_2}}{m_0}, \nonumber \\
c_{fN_{r},fn_{r_2}}=&1.008 + 0.008\frac{\mu_{\lambda,\rho_2}}{m_0},\nonumber\\
c_{0N_{r},0n_{r_2}}=&0.334 - 0.087\frac{\mu_{\lambda,\rho_2}}{m_0},\label{fitcfxnr}
\end{align}
where $m_0=1$ {\gev}. $\mu_{\lambda}$ and $\mu_{\rho_2}$ are the reduced masses defined in Eq. (\ref{pa2qQ}). The relations in Eq. (\ref{fitcfxnr}) are obtained by fitting the mesons, baryons, and tetraquarks.
The {\rt} parameters for the $\rho_1$-mode are calculated by the relations given in Eqs. (\ref{apprelA1}), (\ref{apprelB1}), (\ref{apprelC1}), and (\ref{apprelD1}).
The {\rt} parameters for the $\sigma_1$- and $\sigma_2$-modes are listed in Table \ref{tab:parmv}.

It should be noted that parameters can be quite different in different models: $m_u=0.06$ {\gev} and $\sigma=0.1936$ {\gev}$^2$ in Ref. \cite{Sonnenschein:2018fph}, $m_u=0.30$ {\gev} and $\sigma=0.308$ {\gev}$^2$ in Ref. \cite{Olsson:1994cv}, $m_u=0.33$ {\gev} and $\sigma=0.18$ {\gev}$^2$ in Ref. \cite{Ebert:2009ub}, $m_u=0.22$ {\gev} and $\sigma=0.18$ {\gev}$^2$ in Ref. \cite{Godfrey:1985xj}, and $m_u=0.30$ {\gev} and $\sigma=0.142$ {\gev}$^2$ in Ref. \cite{Veseli:1996gy}. Nevertheless, these distinct parameter values can still produce consistent physical results. Similarly, the parameters adopted in Table \ref{tab:parmv} have been used to estimate masses of mesons, diquarks, baryons, tetraquarks, and pentaquarks \cite{Chen:2023cws,Chen:2023web,Xie:2024dfe,Xie:2024lfo,
Chen:2025fyh,Liu:2026btq}, yielding results consistent with other theoretical predictions. Accordingly, the crude mass estimates for hexaquark ${(\bar{u}(cc))(b(\bar{b}\bar{b}))}$, obtained using the {\rt} relations together with these parameters, are expected to agree with other theoretical predictions. To
our knowledge, no theoretical investigations on the mass spectra of hexaquark ${(\bar{u}(cc))(b(\bar{b}\bar{b}))}$ have been reported to date. Future calculations based on other theoretical approaches can be utilized to further verify our present results.

\begin{table}[!phtb]
\caption{The values of parameters \cite{Feng:2023txx,Faustov:2021hjs,Ebert:2002ig}.}  \label{tab:parmv}
\centering
\begin{tabular*}{0.47\textwidth}{@{\extracolsep{\fill}}cl@{}}
\hline\hline
          & $m_{u}=0.33\; {\gev}$, \; $m_b=4.88\; {\gev}$  \\
          & $m_c=1.55\; {\gev}$, \; $\sigma=0.18\; {\gev^2}$,\; $C=-0.3\; {\gev}$ \\
          & $c_{fn_{r}}=1$,\; $c_{fl}=1.17$\\
$(bb)$    & $c_{0n_{r}}(1^3s_1)=0.01$,\;  $c_{0n_{l}}(1^3s_1)=0.001$\\
$(cc)$    & $c_{0n_{r}}(1^3s_1)=0.205$, \; $c_{0n_{l}}(1^3s_1)=0.337$\\
\hline
\hline
\end{tabular*}
\end{table}

\begin{table}[!phtb]
\caption{The spin-averaged masses of the radially and orbitally $\lambda$-excited states of the {\hxq} (in {\gev}). The notation defined in Eq. (\ref{tetnot}) is rewritten as $|n_1^{2s_1+1}l_{1j_1},n_2^{2s_2+1}l_{2j_2},n_3^{2s_3+1}l_{3j_3},
n_4^{2s_4+1}l_{4j_4},N^{2S+1}L_{J}\rangle$. Specifically, $|n_1l_{1},
n_2l_{2},n_3^{2s_3+1}l_{3j_3},n_4^{2s_4+1}l_{4j_4},NL\rangle$ represents the spin-averaged states. Equations (\ref{t2q}) and (\ref{pa2qQ}) or Eqs. (\ref{combrt}) and (\ref{pa2qQ}) are used in the calculation.}  \label{tab:lambda}
\centering
\begin{tabular*}{0.480\textwidth}{@{\extracolsep{\fill}}cc@{}}
\hline\hline
$|n_1l_{1},n_2l_{2},n_3^{2s_3+1}l_{3j_3},n_4^{2s_4+1}l_{4j_4},NL\rangle$  &  Mass \\
\hline
  $|1s,1s,1^3s_1, 1^3s_1, 1S\rangle$  &17.86  \\
  $|1s,1s,1^3s_1, 1^3s_1, 2S\rangle$  &18.31\\
  $|1s,1s,1^3s_1, 1^3s_1, 3S\rangle$  &18.60\\
  $|1s,1s,1^3s_1, 1^3s_1, 4S\rangle$  &18.85\\
   $|1s,1s,1^3s_1, 1^3s_1, 5S\rangle$  &19.08\\
     $|1s,1s,1^3s_1, 1^3s_1, 1S\rangle$  &17.87      \\
     $|1s,1s,1^3s_1, 1^3s_1, 1P\rangle$  &18.19      \\
     $|1s,1s,1^3s_1, 1^3s_1, 1D\rangle$  &18.41      \\
     $|1s,1s,1^3s_1, 1^3s_1, 1F\rangle$  &18.59      \\
     $|1s,1s,1^3s_1, 1^3s_1, 1G\rangle$  &18.76      \\
\hline\hline
\end{tabular*}
\end{table}

\begin{figure}[!phtb]
\centering
\subfigure[]{\label{subfigure:fa}\includegraphics[scale=0.55]{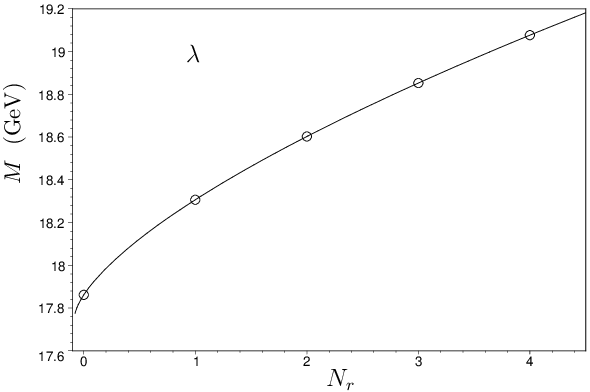}}
\subfigure[]{\label{subfigure:fa}\includegraphics[scale=0.55]{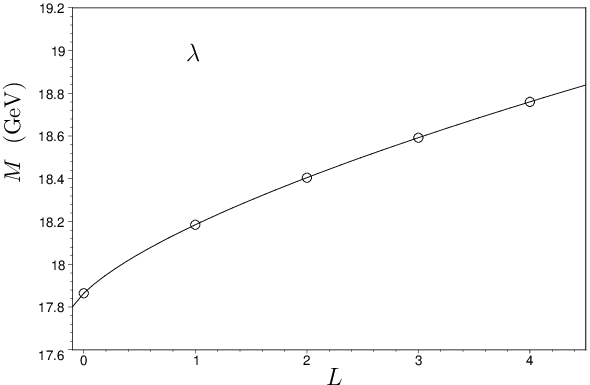}}
\caption{The $\lambda$-{\trs} for the {\hxq}. $N_{r}$ and $L$ are the radial and orbital quantum numbers for the $\lambda$-mode, respectively. Circles represent the predicted data listed in Table \ref{tab:lambda}. The black lines correspond to the fitted formulas which are listed in Table \ref{tab:fitformulas}. }\label{fig:lambda}
\end{figure}

\begin{table}[!phtb]
\caption{The fitted formulas of the $\sigma$-, $\rho$-, and $\lambda$-{\rts} for {\hxq}. }  \label{tab:fitformulas}
\centering
\begin{tabular*}{0.48\textwidth}{@{\extracolsep{\fill}}cl@{}}
\hline\hline
 $\lambda$   & $M=17.767+0.513(0.080+N_r)^{2/3}$  \\
             & $M=17.767+0.386(0.128+L)^{2/3}$ \\
 $\rho_1$    & $M=17.521+0.700\sqrt{0.236+n_{r_1}}$ \\
             & $M=17.515+0.557\sqrt{0.391+{l_1}}$ \\
 $\rho_2$    & $M=17.818+0.312(0.053+n_{r_2})^{2/3}$ \\
             & $M=17.818+0.235(0.084+l_2)^{2/3}$ \\
 $\sigma_1$  & $M=17.687+0.433(0.263+n_{r_3})^{2/3}$ \\
             & $M=17.7+0.338(0.381+l_3)^{2/3}$ \\
 $\sigma_2$  & $M=17.847+0.330(0.010+n_{r_4})^{2/3}$ \\
             & $M=17.847+0.259(0.001+l_4)^{2/3}$ \\
\hline\hline
\end{tabular*}
\end{table}

\subsection{$\lambda$-{\trs} for the {\hxq}}\label{subsec:rts}

The $\lambda$-{\trs} are the simplest among five series of hexaquark {\rts}. Using Eqs. (\ref{pa2qQ}), (\ref{combrt}), (\ref{fitcfxnr}), (\ref{apprelA1}), (\ref{apprelB1}), (\ref{apprelC1}), (\ref{apprelD1}), and data in Table \ref{tab:parmv}, we have the complete form of the $\lambda$-{\trs} and crudely estimate the masses of $\lambda$-excited stats. The calculated masses are listed in Table \ref{tab:lambda}.

When discussing the $\lambda$-{\trs}, other modes are fixed to their radial ground state. The masses of the triquark and antitriquark are known, and all variables except $N_r$ and $L$ are determined; therefore, the $\lambda$-{\trs} become the simplest among five series of hexaquark {\rts}.
The triquark $(\bar{u}(cc))$ and antitriquark $(b(\bar{b}\bar{b}))$ are heavy constituents, therefore, the {\hxq} is a heavy-heavy system. Consequently, the $\lambda$-{\trs} behaves like $M{\sim}x_{\lambda}^{2/3}$ (where $x_{\lambda}=N_r,\,L$) according to Eq. (\ref{massfinal}), as illustrated by the fitted formulas listed in Table \ref{tab:fitformulas}.

In Fig. \ref{fig:lambda}, the circles denote the data calculated using the complete forms of the $\lambda$-{\trs}. The black lines represent the fitted formulas, which are listed in Table \ref{tab:fitformulas}. These fitted formulas are obtained by fitting the values calculated from the complete forms of the $\lambda$-{\trs}. For $\lambda$-{\trs}, these fitted formulas are identical to their complete forms.

\subsection{$\rho_1$- and $\rho_2$-{\trs} for the {\hxq}}\label{subsec:rts}
\begin{table*}[!phtb]
\caption{Same as Table \ref{tab:lambda} except for the $\rho_1$- and $\rho_2$-excited states. }  \label{tab:rho}
\centering
\begin{tabular*}{1.0\textwidth}{@{\extracolsep{\fill}}cccc@{}}
\hline\hline
$|n_1l_{1},n_2l_{2},n_3^{2s_3+1}l_{3j_3},n_4^{2s_4+1}l_{4j_4},NL\rangle$  &  Mass &  $|n_1l_{1},
n_2l_{2},n_3^{2s_3+1}l_{3j_3},n_4^{2s_4+1}l_{4j_4},NL\rangle$ & Mass  \\
\hline
 $|1s,1s,1^3s_1, 1^3s_1, 1S\rangle$  &17.86
    & $|1s,1s,1^3s_1, 1^3s_1, 1S\rangle$  &17.86      \\
 $|2s,1s,1^3s_1, 1^3s_1, 1S\rangle$  &18.30
    & $|1s,2s,1^3s_1, 1^3s_1, 1S\rangle$  &18.14      \\
 $|3s,1s,1^3s_1, 1^3s_1, 1S\rangle$  &18.57
    & $|1s,3s,1^3s_1, 1^3s_1, 1S\rangle$  &18.32      \\
 $|4s,1s,1^3s_1, 1^3s_1, 1S\rangle$  &18.78
    & $|1s,4s,1^3s_1, 1^3s_1, 1S\rangle$  &18.48      \\
 $|5s,1s,1^3s_1, 1^3s_1, 1S\rangle$  &18.96
    & $|1s,5s,1^3s_1, 1^3s_1, 1S\rangle$  &18.61      \\
\hline
 $|1s,1s,1^3s_1, 1^3s_1, 1S\rangle$  &17.86
    & $|1s,1s,1^3s_1, 1^3s_1, 1S\rangle$  &17.86      \\
 $|1p,1s,1^3s_1, 1^3s_1, 1S\rangle$  &18.17
    & $|1s,1p,1^3s_1, 1^3s_1, 1S\rangle$  &18.07      \\
 $|1d,1s,1^3s_1, 1^3s_1, 1S\rangle$  &18.38
    & $|1s,1d,1^3s_1, 1^3s_1, 1S\rangle$  &18.20      \\
 $|1f,1s,1^3s_1, 1^3s_1, 1S\rangle$  &18.54
    & $|1s,1f,1^3s_1, 1^3s_1, 1S\rangle$  &18.32      \\
 $|1g,1s,1^3s_1, 1^3s_1, 1S\rangle$  &18.68
    & $|1s,1g,1^3s_1, 1^3s_1, 1S\rangle$  &18.42      \\
\hline\hline
\end{tabular*}
\end{table*}

\begin{figure*}[!phtb]
\centering
\subfigure[]{\label{subfigure:rfa}\includegraphics[scale=0.41]{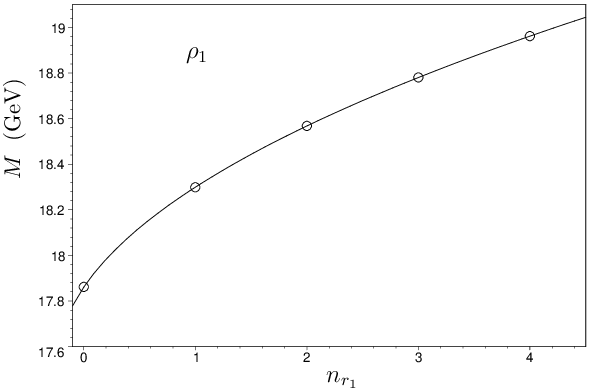}}
\subfigure[]{\label{subfigure:rfb}\includegraphics[scale=0.41]{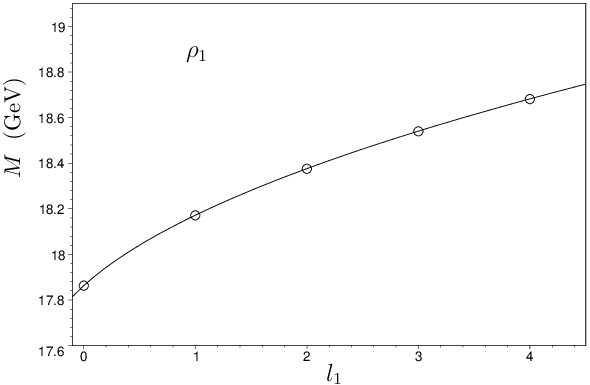}}
\subfigure[]{\label{subfigure:rfc}\includegraphics[scale=0.41]{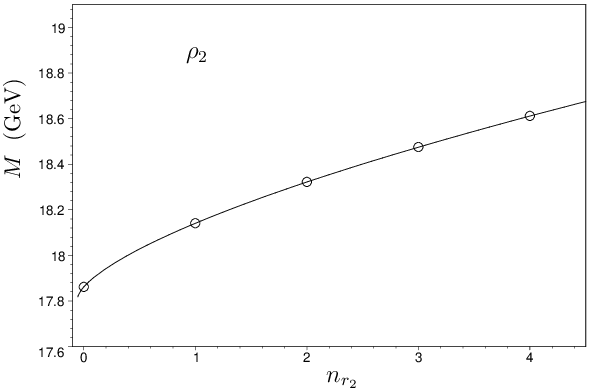}}
\subfigure[]{\label{subfigure:rfd}\includegraphics[scale=0.41]{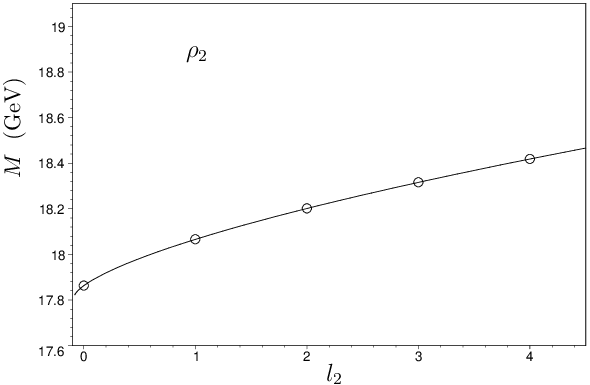}}
\caption{The $\rho_1$- and $\rho_2$-{\trs} for the {\hxq}. $n_{r_1}$ and ${l_1}$ are the radial and orbital quantum numbers for the $\rho_1$-mode, respectively. $n_{r_2}$ and ${l_2}$ are the radial and orbital quantum numbers for the $\rho_2$-mode, respectively.  Circles represent the predicted data listed in Table \ref{tab:rho}. The black lines correspond to the fitted formulas, obtained by fitting the calculated data in Table \ref{tab:rho}. These formulas are listed in Table \ref{tab:fitformulas}. }\label{fig:rho}
\end{figure*}

In this subsection, both the $\rho_1$- and $\rho_2$-modes are investigated. When discussing the $\rho_1$-mode, all other modes are fixed at their radial ground states, with the adopted parameters matching those of the radial ground state. By the similar procedure, the $\rho_2$-mode can be discussed.

In contrast to the hexaquark $\lambda$-{\trs} and the {\rts} for light mesons, which have very simple forms, the $\rho_1$- and $\rho_2$-{\trs} are more complex. For instance, the complete form of the radial $\rho_1$-{\tr} reads
\begin{widetext}
\begin{align}\label{rhoart}
M=&17.3988+ 0.741618 \sqrt{  0.246694+ n_{r_1}} +0.292339
  \left(0.334 -  \frac{1.25273 \left(3.29958+ 0.741618 \sqrt{0.246694+ {n_{r_1}}}\right)}{  17.6988+ 0.741618 \sqrt{0.246694+ {n_{r_1}}}}\right)^{2/3} \nonumber\\
  &    \times\left(1.008+  \frac{ 0.115194 \left(3.29958+ 0.741618\sqrt{0.246694+ n_{r_1}}\right)} {17.6988+   0.741618 \sqrt{0.246694+ n_{r_1}}}\right)
\left(\frac{3.29958+ 0.741618 \sqrt{0.246694+ n_{r_1}}}
  { 17.6988+ 0.741618 \sqrt{0.246694+ n_{r_1}}}\right)^{-1/3}.
\end{align}
\end{widetext}
By fitting the data calculated from the complete form of the $\rho_1$-{\tr} [Eq. (\ref{rhoart})], we obtain a simple fitted formula (listed in Table \ref{tab:fitformulas})
\bea\label{rhoartfit}
M=17.521+0.700\sqrt{0.236+n_{r_1}}.
\eea
Since (\ref{rhoart}) can be well approximated by (\ref{rhoartfit}), they have the same behavior, $M{\sim}\sqrt{n_{r_1}}$.
The radial triquark $(\bar{u}(cc))$ {\rt} reads, from Eqs. (\ref{t2q}) and (\ref{pa2qQ}),
\bea\label{trrt}
M=3.29958+0.741618 \sqrt{0.246694+n_{r_1}}.
\eea
Comparing Eqs. (\ref{rhoart}) and (\ref{trrt}) reveals that the radial $\rho_1$-{\tr} [Eq. (\ref{rhoart})] for the {\hxq} is distinct from the radial triquark $(\bar{u}(cc))$ {\rt}. Nevertheless, both have the same Regge trajectory behavior, $M{\sim}\sqrt{n_{r_1}}$, see Eqs. (\ref{rhoart}), (\ref{rhoartfit}), and (\ref{trrt}).

Using Eq. (\ref{rhoart}), masses of the radially $\rho_1$-excited states can be crudely estimated, which are listed in Table \ref{tab:rho}.
The radial $\rho_1$-{\tr} is shown in Fig. \ref{subfigure:rfa}. The circles represent data calculated from the complete form of the $\rho_1$-{\tr} [Eq. (\ref{rhoart})]. The black line represents the fitted formula [in Eq. (\ref{rhoartfit})], which is listed in Table \ref{tab:fitformulas}.

By a similar procedure, the orbital $\rho_1$-{\tr} and the $\rho_2$-{\rts} can be obtained. Results are listed in Table \ref{tab:rho} and shown in Fig. \ref{fig:rho}. The orbital $\rho_1$-{\tr} behaves as $M{\sim}\sqrt{l_1}$, whereas the $\rho_2$-{\trs} follow $M{\sim}n_{r_2}^{2/3},\,l_2^{2/3}$.

Inspection of Eqs. (\ref{pa2qQ}), (\ref{combrt}), (\ref{fitcfxnr}), (\ref{apprelA1}), (\ref{apprelB1}), (\ref{apprelC1}), (\ref{apprelD1}), and (\ref{rhoart}) reveals that the $\rho_1$- and $\rho_2$-{\trs} cannot be directly constructed by mimicking the meson {\rts}, because mesons have no substructures whereas hexaquarks have structures and substructures.
If the structures and substructures are unavailable, the $\rho_1$-, and $\rho_2$-trajectories can only be determined by fitting either existing theoretical predictions or future experimental measurements.

\subsection{$\sigma_1$- and $\sigma_2$-{\trs} for the {\hxq}}\label{subsec:rts}

\begin{table*}[!phtb]
\caption{Same as Table \ref{tab:lambda} except for the $\sigma_1$- and $\sigma_2$-excited states of the {\hxq}. $(\times)$ denotes the nonexistent states.}  \label{tab:sigma}
\centering
\begin{tabular*}{1.0\textwidth}{@{\extracolsep{\fill}}cccc@{}}
\hline\hline
$|n_1l_{1},n_2l_{2},n_3^{2s_3+1}l_{3j_3},n_4^{2s_4+1}l_{4j_4},NL\rangle$  &  Mass &  $|n_1l_{1},
n_2l_{2},n_3^{2s_3+1}l_{3j_3},n_4^{2s_4+1}l_{4j_4},NL\rangle$ & Mass  \\
\hline
 $|1s,1s,1^3s_1, 1^3s_1, 1S\rangle$  & 17.86
    & $|1s,1s,1^3s_1, 1^3s_1, 1S\rangle$  &   17.86   \\
 $|1s,1s,2^3s_1, 1^3s_1, 1S\rangle$  & 18.19
    & $|1s,1s,1^3s_1, 2^3s_1, 1S\rangle$  &   18.18   \\
 $|1s,1s,3^3s_1, 1^3s_1, 1S\rangle$  &  18.43
    & $|1s,1s,1^3s_1, 3^3s_1, 1S\rangle$  &   18.37  \\
 $|1s,1s,4^3s_1, 1^3s_1, 1S\rangle$  &   18.64
    & $|1s,1s,1^3s_1, 4^3s_1, 1S\rangle$  &    18.53  \\
 $|1s,1s,5^3s_1, 1^3s_1, 1S\rangle$  &   18.83
    & $|1s,1s,1^3s_1, 5^3s_1, 1S\rangle$  &    18.68  \\
\hline
 $|1s,1s,1^3s_1, 1^3s_1, 1S\rangle$  &17.88
    & $|1s,1s,1^3s_1, 1^3s_1, 1S\rangle$  & 17.85     \\
 $|1s,1s,1^3p_2, 1^3s_1, 1S\rangle$ $(\times)$ &18.12
    & $|1s,1s,1^3s_1, 1^3p_2, 1S\rangle$ $(\times)$ & 18.11     \\
 $|1s,1s,1^3d_3, 1^3s_1, 1S\rangle$  &18.30
    & $|1s,1s,1^3s_1, 1^3d_3, 1S\rangle$  &   18.26   \\
 $|1s,1s,1^3f_4, 1^3s_1, 1S\rangle$ $(\times)$ &18.46
    & $|1s,1s,1^3s_1, 1^3f_4, 1S\rangle$ $(\times)$  & 18.39     \\
 $|1s,1s,1^3g_5, 1^3s_1, 1S\rangle$  &18.60
    & $|1s,1s,1^3s_1, 1^3g_5, 1S\rangle$  & 18.50     \\
\hline\hline
\end{tabular*}
\end{table*}

\begin{figure*}[!phtb]
\centering
\subfigure[]{\label{subfigure:cfa}\includegraphics[scale=0.41]{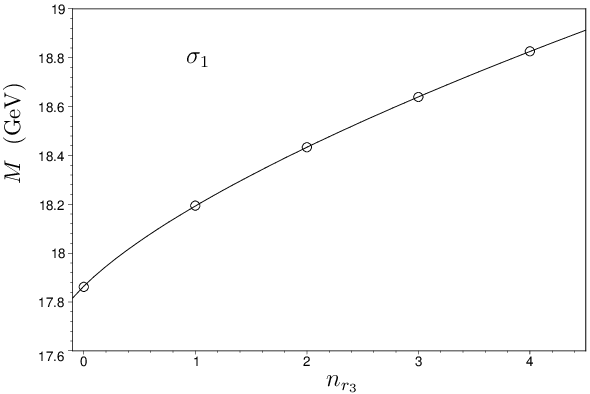}}
\subfigure[]{\label{subfigure:cfa}\includegraphics[scale=0.41]{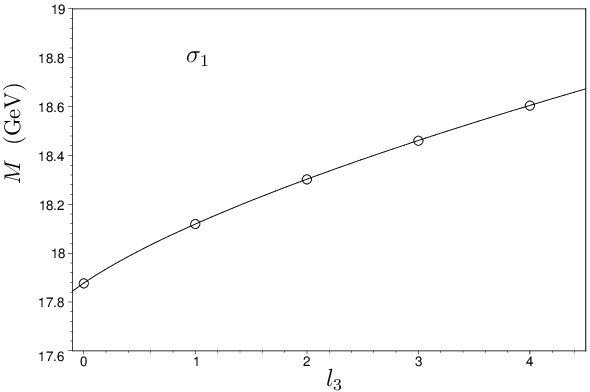}}
\subfigure[]{\label{subfigure:cfa}\includegraphics[scale=0.41]{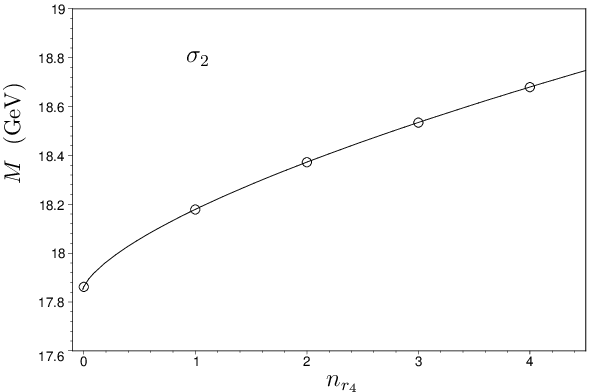}}
\subfigure[]{\label{subfigure:cfa}\includegraphics[scale=0.41]{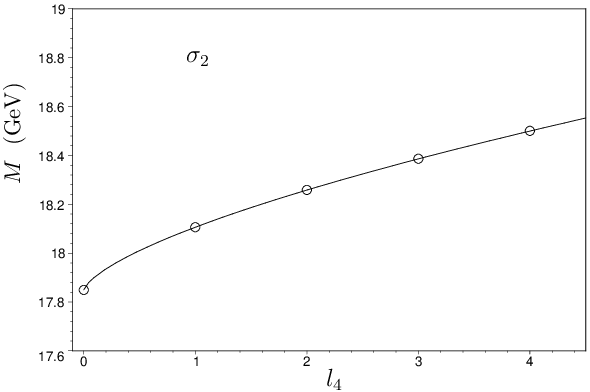}}
\caption{The $\sigma_1$- and $\sigma_2$-{\trs} for the {\hxq}. $n_{r_3}$ and ${l_3}$ are the radial and orbital quantum numbers for the $\sigma_1$-mode, respectively, while $n_{r_4}$ and ${l_4}$ are the corresponding numbers for the $\sigma_2$-mode.  Circles represent the predicted data listed in Table \ref{tab:sigma}. The black lines correspond to the fitted formulas, obtained by fitting the predicted data in Table \ref{tab:sigma}. These formulas are listed in Table \ref{tab:fitformulas}.}\label{fig:sigma}
\end{figure*}

In this subsection, the $\sigma_1$- and $\sigma_2$-{\trs} for the {\hxq} are given using Eqs. (\ref{pa2qQ}), (\ref{combrt}), (\ref{fitcfxnr}), (\ref{apprelA1}), (\ref{apprelB1}), (\ref{apprelC1}), (\ref{apprelD1}), and data in Table \ref{tab:parmv}. These {\trs} are visualized in Fig. \ref{fig:sigma}. Meanwhile, the masses of $\sigma_1$- and $\sigma_2$-excited states are roughly estimated and listed in Table \ref{tab:sigma}. $1^3p_2$ and $1^3f_4$ do not exist for diquarks $(bb)$ and $(cc)$, which are denoted by $\times$ in Table \ref{tab:sigma}.
Similar to the $\rho$-{\trs}, the $\sigma$-{\trs} are constructed based on the intrinsic structures and substructures of hexaquark.
We stress that the $\sigma_1$- and $\sigma_2$-{\trs} are not the diquark and antidiquark {\rts}, respectively. However, they share the same trajectory behaviors, respectively.

Among five series of hexaquark {\rts}, the $\sigma_1$- and $\sigma_2$-{\trs} are the lengthiest. By fitting the data calculated from the complete forms of $\sigma$-{\trs}, the fitted formulas are obtained and listed in Table \ref{tab:fitformulas}. All radial and orbital $\sigma_1$- and $\sigma_2$-{\trs} behave as $M{\sim}x_{\sigma}^{2/3}$ $(x_{\sigma}=n_{r_3},\, n_{r_4},\,l_3,\,l_4)$.

\section{Conclusion}\label{sec:conc}
In this work, we propose {\rt} relations for the quintuply heavy {\hxq} by employing the diquark and triquark {\rts}. Using these newly derived relations, we systematically investigate five series of hexaquark Regge trajectories, namely, the $\lambda$-, $\rho_1$-, $\rho_2$-, $\sigma_1$-, and $\sigma_2$-trajectories. The masses of the $\lambda$-, $\rho_1$-, $\rho_2$-, $\sigma_1$-, and $\sigma_2$-excited states are roughly estimated.

The complete forms of the Regge trajectories for {\hxq} are lengthy and cumbersome. Except for the $\lambda$-{\trs}, the $\rho_1$-, $\rho_2$-, $\sigma_2$-, and $\sigma_2$-{\trs} cannot be constructed by simply mimicking the meson {\rts}, instead, they should be constructed based on the hexaquark structures and substructures.
Otherwise, the $\rho_1$-, $\rho_2$-, $\sigma_1$-, and $\sigma_2$-trajectories can only be determined by fitting either theoretical predictions or experimental measurements. The fundamental relationship between the slopes of the obtained trajectories, constituents' masses, and string tension would become obscure.

We demonstrate that the $\rho_1$-, $\rho_2$-, $\sigma_1$-, and $\sigma_2$-{\trs} are not the triquark, antitriquark, diquark, and antidiquark {\rts}, respectively. However, they share identical trajectory behaviors, respectively.
We show that the lengthy complete forms of the $\rho_1$-, $\rho_2$-, $\sigma_1$-, and $\sigma_2$-trajectories can be well approximated by the simple fitted formulas.
The $\rho_1$-{\trs} behave as $M{\sim}\sqrt{x_{\rho_1}}$ $(x_{\rho_1}=n_{r_1},\,l_1)$, while the $\lambda$-, $\rho_2$-, $\sigma_1$-, and $\sigma_2$-{\trs} all exhibit behavior of $M{\sim}x^{2/3}$, where  $x=N_r,\,L,\,,n_{r_2},\,l_2,\,n_{r_3},\,l_3,\,n_{r_4},\,l_4$.

Reference \cite{Pan:2025gqn} discusses hexaquarks with various quark contents, including $qQQQQQ$. Both ${(\bar{u}(cc))(b(\bar{b}\bar{b}))}$ and $qQQQQQ$ are quintuply heavy hexaquarks. Although there is no sufficient experimental information on these hexaquarks, the results still have a certain guiding significance for future research.





\section*{DATA AVAILABILITY}
There are no publicly available research data or software supporting this manuscript. Requests for further information or data should be sent to the authors.

\appendix
\section{Determination of $c_{fx_{\lambda}}$ and $c_{0x_{\lambda}}$ for the heavy-light systems}\label{sec:appcfx}

In this appendix, we determine the values of $c_{fx_{\lambda}}$ and $c_{0x_{\lambda}}$ for the $\lambda$-modes of the {\dhbs} $(QQ')u$ with $Q,Q'=b,c$.
Equation (\ref{massfinal}) is applied to fit the {\rts} for heavy-light mesons and the $\lambda$-modes of doubly heavy baryons, which are composed of one doubly heavy diquark and one light quark.
The quality of a fit is measured by the quantity $\chi^2$, defined as
\bea\label{appfit}
\chi^2=\frac{1}{N-1}\sum^{N}_{i=1}\left(\frac{M_{fi}-M_{ei}}{M_{ei}}\right)^2,
\eea
where $N$ is the number of points on a trajectory, $M_{fi}$ stands for the fitted value, and $M_{ei}$ is the experimental value or the theoretical value of the $i$-th particle mass. The parameters are determined by minimizing $\chi^2$.

\begin{table}[!phtb]
\caption{Spin-averaged masses of radially excited states for heavy-light mesons and doubly heavy baryons (in ${\gev}$). $\{cc\}_1$, $\{cc\}_2$, $\{bb\}_1$, and $\{bb\}_2$ correspond, respectively, to the ground state of axial vector diquark $\{cc\}$, the first radially excited state of axial vector diquark $\{cc\}$, the ground state of axial vector diquark $\{bb\}$, and the first radially excited state of axial vector diquark $\{bb\}$.}  \label{tab:cfr}
\centering
\begin{tabular*}{0.45\textwidth}{@{\extracolsep{\fill}}cccccc@{}}
\hline\hline
           & $1S$  &  $2S$  &  $3S$  & $4S$  &  $5S$    \\
\hline
$c\bar{u}$  &1.971  &2.608   &3.088   &3.475   &3.815    \\
$b\bar{u}$  &5.313  &5.902   &6.385   &6.785   &7.132    \\
$\{cc\}_1u$     &3.691  &4.356   &   &   &    \\
$\{cc\}_2u$     &3.988  &4.649   &   &   &    \\
$\{bb\}_1u$     &10.225  &10.851   &   &   &    \\
$\{bb\}_2u$     &10.468  &11.082   &   &   &    \\
\hline\hline
\end{tabular*}
\end{table}

\begin{table}[!phtb]
\caption{Same as Table \ref{tab:cfr} except for the orbitally excited states.}  \label{tab:cfo}
\centering
\begin{tabular*}{0.45\textwidth}{@{\extracolsep{\fill}}cccccc@{}}
\hline\hline
            & $1S$  &  $1P$  &  $1D$  & $1F$  &  $1G$    \\
\hline
$c\bar{u}$  &1.971   &2.429  &2.772  &3.145  &3.417   \\
$b\bar{u}$  &5.313   &5.745  &6.106  &6.398  &6.648   \\
$\{cc\}_1u$     &3.691   &4.139  &  &  &   \\
$\{cc\}_2u$     &3.988   &4.480  &  &  &   \\
$\{bb\}_1u$     &10.225   &10.664  &  &  &   \\
$\{bb\}_2u$     &10.468   &10.901  &  &  &   \\
\hline\hline
\end{tabular*}
\end{table}

\begin{table}[!phtb]
\caption{Fitted values of $c_{fx_{\lambda}}$ and $c_{0x_{\lambda}}$. $\{cc\}_1$, $\{cc\}_2$, $\{bb\}_1$, and $\{bb\}_2$ correspond respectively to the ground axial vector diquark $\{cc\}$, the first radially excited axial vector diquark $\{cc\}$, the ground axial vector diquark $\{bb\}$, and the first radially excited axial vector diquark $\{bb\}$.}  \label{tab:fp}
\centering
\begin{tabular*}{0.45\textwidth}{@{\extracolsep{\fill}}ccc@{}}
\hline\hline
          & $(c_{fN_r},\,c_{0N_r})$  & $(c_{fL},\,c_{0L})$ \\

$c\bar{u}$    & $(1.0012,\; 0.126)$   & $(1.0139,\; 0.188)$  \\
$\{cc\}_1u$  & $(1.0135,\; 0.252)$   & $(0.9762,\; 0.426)$ \\
$\{cc\}_2u$  & $(0.9595,\; 0.202)$   & $(0.9814,\; 0.302)$ \\
$b\bar{u}$    & $(0.9880,\; 0.128)$   & $(0.9706,\; 0.216)$ \\
$\{bb\}_1u$  & $(0.9861,\; 0.290)$   & $(0.9778,\; 0.464)$ \\
$\{bb\}_2u$  & $(0.9290,\; 0.244)$   & $(0.9207,\; 0.390)$ \\
\hline\hline
\end{tabular*}
\end{table}

The masses used are listed in Tables \ref{tab:cfr} and \ref{tab:cfo}. For experimentally determined states, the Particle Data Group values are taken from Ref. \cite{pdg2026}. For the undetermined states, the data are from Refs. \cite{Ebert:2002ig,Ebert:2009ua}.
In addition to the masses of bound states, some parameters are provided in Table \ref{tab:parmv}.
The masses of four axial vector diquarks are listed as follows: the ground state $\{cc\}_1$, the first radially excited state $\{cc\}_2$, the ground state $\{bb\}_1$, and the first radially excited state $\{bb\}_2$ have masses of $3.12$ {\gev}, $3.50$ {\gev}, $9.63$ {\gev}, and $9.95$ {\gev}, respectively. The masses of diquarks $(QQ')$ ($Q,Q'=b,c$) are taken from Ref. \cite{Feng:2023txx}, these diquark masses can be calculated using the parameters in Table \ref{tab:parmv} along with Eqs. (\ref{t2q}) and (\ref{pa2qQ}).

\begin{figure*}[!phtb]
\centering
\subfigure[]{\label{subfigure:cfa}\includegraphics[scale=0.4]{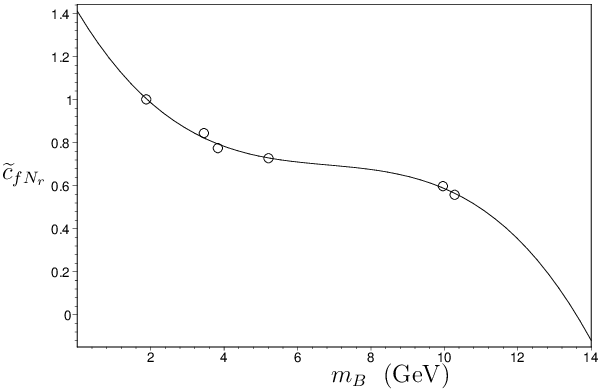}}
\subfigure[]{\label{subfigure:cfa}\includegraphics[scale=0.4]{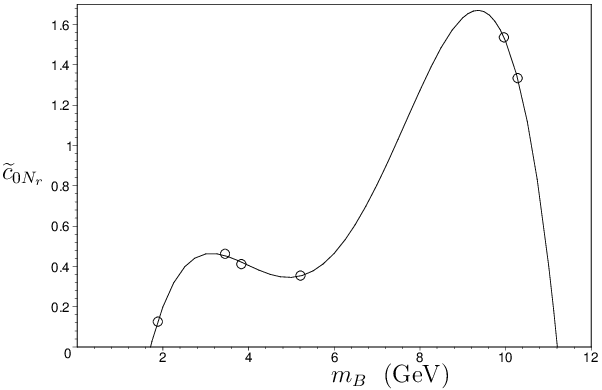}}
\subfigure[]{\label{subfigure:cfa}\includegraphics[scale=0.4]{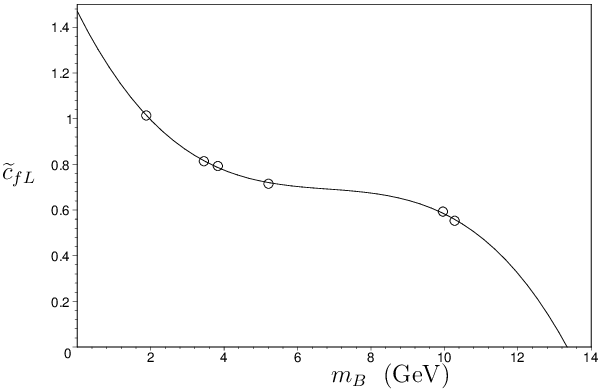}}
\subfigure[]{\label{subfigure:cfa}\includegraphics[scale=0.4]{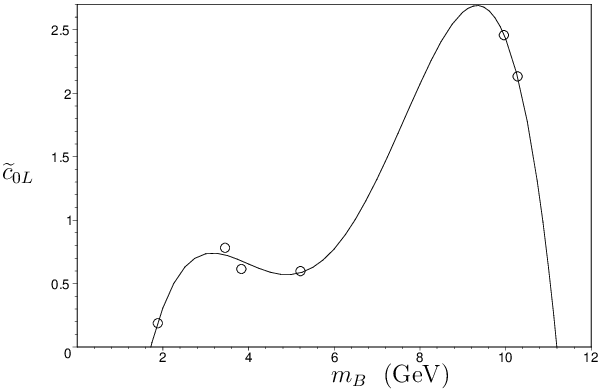}}
\caption{Fitted curves for $\widetilde{c}_{fN_r}$, $\widetilde{c}_{0N_r}$, $\widetilde{c}_{fL}$, and $\widetilde{c}_{0L}$. Circles denote data calculated using values listed in Table \ref{tab:fp} and relations linking $\widetilde{c}_{fN_r}$, $\widetilde{c}_{0N_r}$, and $\widetilde{c}_{fL}$, $\widetilde{c}_{0L}$ to ${c}_{fN_r}$, ${c}_{0N_r}$, ${c}_{fL}$, and ${c}_{0L}$, respectively. The lines represent the formulas given in Eqs. (\ref{apprelA}), (\ref{apprelB}), (\ref{apprelC}), and (\ref{apprelD}), respectively. }\label{fig:fitpara}
\end{figure*}

Using Eq. (\ref{massfinal}) with $\sigma'=C$, the coefficients for heavy-light systems given in Table \ref{tab:eparam}, Eq. (\ref{appfit}), and data listed in Tables \ref{tab:cfr} and \ref{tab:cfo}, we obtain the fitted values, as summarized in Table \ref{tab:fp}. Using the fitted values in Table \ref{tab:fp}, we obtain the fitted relation \footnote{In practice, we find that fitting the scaled function $\widetilde{c}_{fN_r}$ yields better results than fitting directly ${c}_{fN_r}$. This analogous behavior holds for all subsequent fittings. }
\begin{align}\label{apprelA}
\widetilde{c}_{fN_r}=&1.4115- 0.2835 \frac{m_B}{m_0} + 0.03927 \left(\frac{m_B}{m_0}\right)^2 \nonumber\\
               &- 0.00192 \left(\frac{m_B}{m_0}\right)^3,
\end{align}
where $m_0=1$ {\gev} and $m_B=m_u+m_{H}$. Here, $m_H$ is the heavy quark mass for heavy-light meson, whereas it represents the mass of the doubly heavy diquark for doubly heavy baryons. ${c}_{fN_r}$ is recovered via ${c}_{fN_r}=\widetilde{c}_{fN_r}m_B^{0.3}(m_c+m_u)^{-0.3}$, i.e.,
\begin{align}\label{apprelA1}
{c}_{fN_r}=&\left(\frac{m_B}{m_c+m_u}\right)^{0.3}\left(1.4115- 0.2835 \frac{m_B}{m_0} \right. \nonumber\\
&   \left. + 0.03927 \left(\frac{m_B}{m_0}\right)^2 - 0.00192 \left(\frac{m_B}{m_0}\right)^3\right).
\end{align}
By fitting the scaled function $\widetilde{c}_{0N_r}$, we have
\begin{align}\label{apprelB}
\widetilde{c}_{0N_r}=& -3.2445 + 3.3231 \frac{m_B}{m_0} - 1.0442 \left(\frac{m_B}{m_0}\right)^2 \nonumber\\
   &+ 0.1333 \left(\frac{m_B}{m_0}\right)^3 -  0.005737 \left(\frac{m_B}{m_0}\right)^4,
\end{align}
where $\widetilde{c}_{0N_r}={c}_{0N_r}m_B(m_c+m_u)^{-1}$. Then, we have
\begin{align}\label{apprelB1}
{c}_{0N_r}=&\frac{m_c+m_u}{m_B}\left(-3.2445 + 3.3231 \frac{m_B}{m_0}  \right. \nonumber\\
&   \left. - 1.0442 \left(\frac{m_B}{m_0}\right)^2+ 0.1333 \left(\frac{m_B}{m_0}\right)^3 \right.\nonumber\\
&\left. -  0.005737 \left(\frac{m_B}{m_0}\right)^4\right).
\end{align}
Performing weighted fitting gives
\begin{align}\label{apprelC}
\widetilde{c}_{fL}=&1.4697- 0.3188 \frac{m_B}{m_0} +
 0.04506 \left(\frac{m_B}{m_0}\right)^2 \nonumber\\
 &- 0.002205 \left(\frac{m_B}{m_0}\right)^3,
\end{align}
where $\widetilde{c}_{fL}={c}_{fL}m_B^{-0.3}(m_c+m_u)^{0.3}$. Then, we have
\begin{align}\label{apprelC1}
{c}_{fL}=&\left(\frac{m_B}{m_c+m_u}\right)^{0.3}\left(1.4697- 0.3188 \frac{m_B}{m_0} \right. \nonumber\\
&   \left. +  0.04506 \left(\frac{m_B}{m_0}\right)^2 - 0.002205 \left(\frac{m_B}{m_0}\right)^3\right).
\end{align}
Fitting the scaled function $\widetilde{c}_{0L}$ gives
\begin{align}\label{apprelD}
\widetilde{c}_{0L}=& -5.1878 + 5.2976 \frac{m_B}{m_0} - 1.6633 \left(\frac{m_B}{m_0}\right)^2 \nonumber\\
    & + 0.2127 \left(\frac{m_B}{m_0}\right)^3 - 0.009171 \left(\frac{m_B}{m_0}\right)^4,
\end{align}
where $\widetilde{c}_{0L}={c}_{0L}m_B(m_c+m_u)^{-1}$. Then, we have
\begin{align}\label{apprelD1}
{c}_{0L}=&\frac{m_c+m_u}{m_B}\left(-5.1878 + 5.2976 \frac{m_B}{m_0} - 1.6633 \left(\frac{m_B}{m_0}\right)^2 \right. \nonumber\\
&   \left. + 0.2127 \left(\frac{m_B}{m_0}\right)^3 - 0.009171 \left(\frac{m_B}{m_0}\right)^4\right).
\end{align}
The scaled fittings are shown in Fig. \ref{fig:fitpara}.

\end{document}